# The Day-night Variation of Cosmic Rays Intensity at Sea Level Under the Influence of Meteorological Fronts and Troughs

H.M.Mok[*] and K.M.Cheng[**]

Radiation Health Unit, 3/F., Saiwanho Health Centre,
HKSAR Government, 28 Tai Hong St.,
Saiwanho, Hong Kong, China
e-mail : [*] jhmmok@netvigator.com
[**] rhuichk@netvigator.com

INTRODUCTION

The cosmic rays (CR) at sea level comprise particles from the extensive air showers (EAS). Their major components are muons, which are the decay products of pions or kaons from high altitudes. Despite a lifetime of only about 2 µs, muons can penetrate the atmosphere down to the sea level before they decay because of the relativistic time dilation effect. When muons travel through the atmosphere, they lose their energy by the process of ionisation. The bremsstrahlung loss of muons is relatively insignificant because of their large mass. Electrons, being a decay product of muons, also appear in the cosmic rays at sea level. However, their penetration power is low because relativistic electrons undergo large bremsstrahlung energy losses. Neutrons also contribute to the sea level cosmic rays as a hadronic component. Low energy neutrons (about 10 MeV or less) are mainly produced by evaporation from the highly excited nuclei. High energy neutrons are produced by the knock-on collision or charge exchange reactions of high energy protons. References (1,2) provide concise introductions on the formation and properties of CR.

The intensity of CR at sea level is not a constant. Many factors, including the barometric effect, solar cycle modulation effect, solar flares effect, geomagnetic effect, etc., may affect its value (3). For muons, because of decay effects, the sea level intensity is related to the height of production. Blackett (4) was probably the first pioneer to explain these decay effects. He pointed out that the pressure level in the atmosphere, where most of the muons were formed (now known to be at about 100 mbar), would be higher as the atmospheric temperature increased. Consequently more muons would have decayed before reaching the ground and thus resulting in a decrease of ionisation. This muon decay effect (or say temperature effect) explains the seasonal variations of the CR intensity at sea level. Forbush (5) also reported that there was random behaviour in the CR intensity when the observing station was under the influence of a meteorological frontal system.

In this study, we observed the day-night variation of the sea level CR intensity using two independent sets of simple counter telescopes each performing two 5-hours counting during the day and the night. The variation patterns were carefully observed whenever our station was under the influence of a meteorological front or trough. Our observation showed that the pattern of variation was closely related to the atmospheric disturbance. Normally, the day counts were a few percents lower than that of the night counts. This could have been explained by the day-night temperature effect of muon decay as mentioned above. But a reversal occurred when our observation station was being affected by a meteorological front or trough. Our further investigation showed that the pattern of variation could have been negatively correlated to the altitude of the $0^oC$ level of the atmosphere. It appeared that the CR intensity at sea level and atmospheric stability had an inverse correlation rather than the random effect described by Forbush. We also observed that solar activities, such as solar flares (especially the X-Class flares) or magnetic storms, could mask this correction. We shall discuss this observation in this paper

MEASUREMENT METHOD

We constructed a simple coincidence counter telescope using two Geiger-Muller (GM) tubes, each of length 32 cm and diameter 3 cm. The output of each GM tube was first connected to a Timing Single Channel Analyser (TSCA) and both output timing signals were fed to a coincidence module and then to an acquisition board for counting by a computer. The counting interval was arbitrarily set at 30 s. Such a counter configuration would mainly measure the charged shower particles (primarily muons). We used two sets of counter telescopes simultaneously to check for the consistency of the data and to exclude any pattern that was caused by random fluctuation. In order to minimize the contribution to variation of CR counts caused by atmospheric tidal effect on air pressure, we measured the day counts and the night counts at 10:00 to 15:00 (local time) and at 17:00 to 22:00 (local time) respectively. We collected data for over 10 months at an observing station located at $114.25^oE$, $22.5^oN$.

Weather information, including weather charts at different altitudes, at 05:00, 08:00, 14:00 and 20:00 (local time) and meteorological data, including the altitude of the $0^oC$ level, were obtained from the web site of the Hong Kong Observatory (HKO) (6). We also obtained the daily solar information from the National Oceanic and Atmospheric Administration (NOAA) through their web-site for discriminating the solar effects from the atmospheric effects. The data of X-ray flux measured by the GOES 8 and 10 satellites of NOAA (7) provided information of the solar activities and the solar flare events while the spaceweather report of NOAA (8) provided the magnetic field and magnetic storm information.

OBSERVATION RESULTS

We present below the results of our observation in the 5 months June to October 1999. We are yet analysing the later data. Figure 1 and 2 shows the plots of relative CR counts, which are the ratio of the CR counts to the average CR count in 5 months, against the ground level atmospheric pressures. In order to minimize the influence of the temperature effect, the day counts and the night counts are presented on separate graphs. Furthermore only the data from June to August 1999 are shown to avoid the seasonal effect. While the points are quite scattered, a generally decreasing trend due to the well-known barometric effect can be observed. However, except in the case of a sudden drop of atmospheric pressure, the day-to-day variation of CR counts due to barometric effect is small (estimated from the figures to be about 0.2%/mb), so that the major variation of CR counts is not due to the barometric effect.

Our observation results show that in the absence of a meteorological front, trough or solar effects, the day count of the CR is usually smaller than the night count by 3-4 %. As discussed in the introduction, this can be attributed to the day-night temperature effect of muon decay. In figure 3-7, two increments on the x-axis represent the data for one day. The first increment of each day is the counting collected during the day and the second is the counting collected during the night.

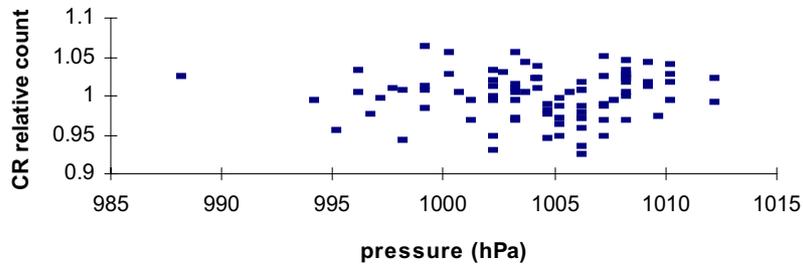

Figure 1. The plot of the day-time relative CR counts with the ground level atmospheric pressure.

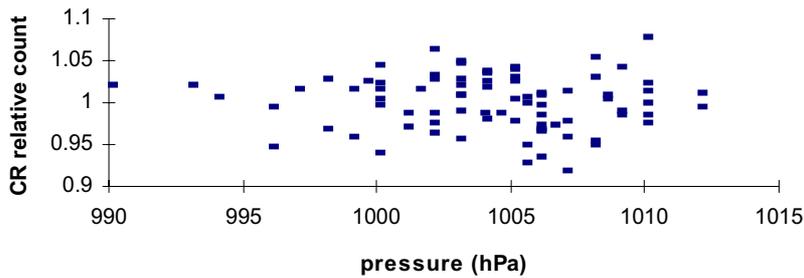

Figure 2. The plot of the night-time relative CR counts with the ground level atmospheric pressure

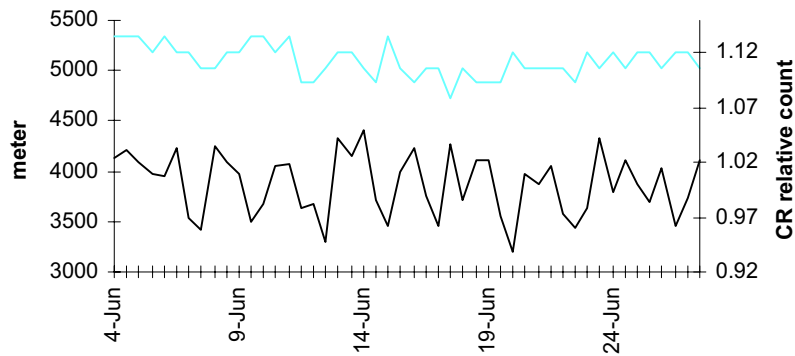

Figure 3. The plot of the day-night variation of CR intensity in June 99 with the $0^{o}C$ level height. The upper curve represents the $0^{o}C$ level height in metre while the lower curve represents the relative CR counts.

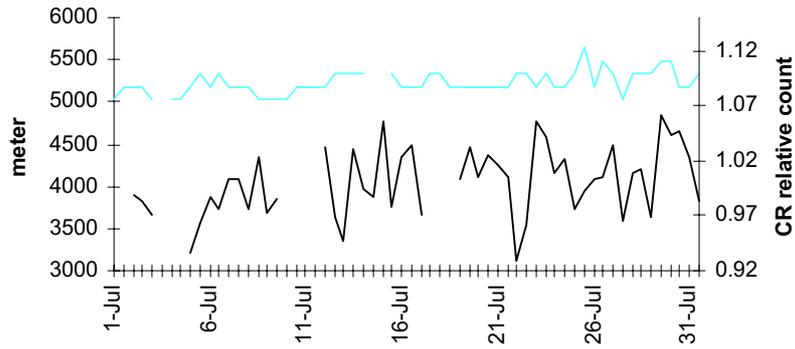

Figure 4. The plot of the day-night variation of CR intensity in July 99 with the $0^\circ$C level height. The upper curve represents the $0^\circ$C level height in metre while the lower curve represents the relative CR counts.

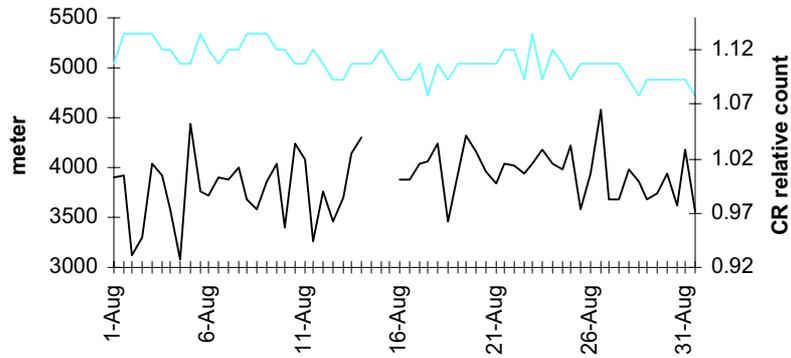

Figure 5. The plot of the day-night variation of CR intensity in August 99 with the $0^\circ$C level height. The upper curve represents the $0^\circ$C level height in metre while the lower curve represents the relative CR counts.

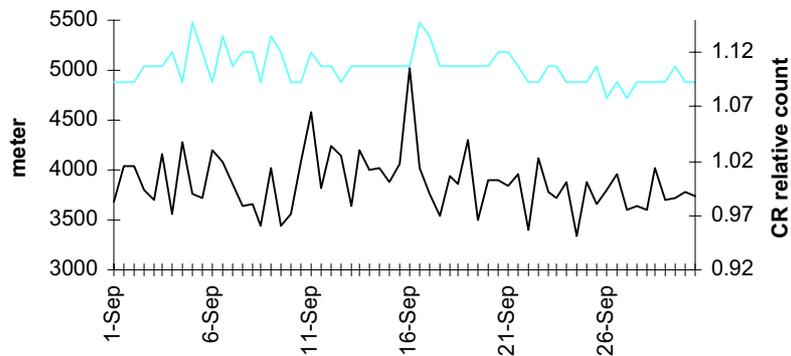

Figure 6. The plot of the day-night variation of CR intensity in September 99 with the $0^\circ$C level height. The upper curve represents the $0^\circ$C level height in metre while the lower curve represents the relative CR counts.

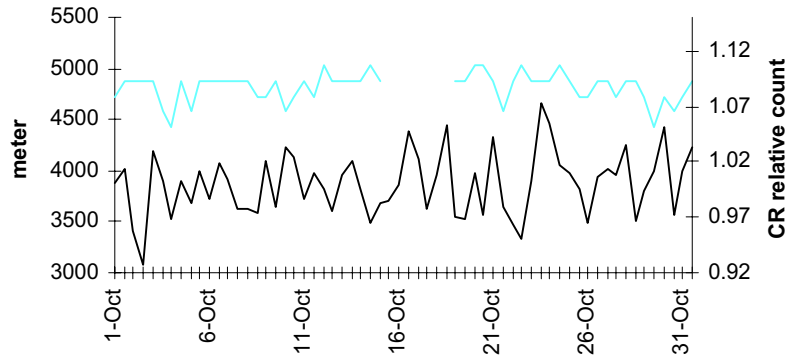

Figure 7. The plot of the day-night variation of CR intensity in October 99 with the $0^{o}C$ level height. The upper curve represents the $0^{o}C$ level height in metre while the lower curve represents the relative CR counts.

We observed that when a front or a trough develops and moves near our station, a reversal pattern occurs. The graphs would show negative slope at the first increment of a date. For example, from 5 to 9 June, the reversal was probably related to the formation a typhoon (Typhoon 岼aggie"). The reversals on 11 and 16 June were probably related to frontal formation. From 12 to 17 July, the reversals were probably related to the frequent trough activities. The reversal on 23 July was probably caused by the formation of a tropical cyclone and so on. Repeated observations of such reversal patterns at changes of atmospheric stability indicated that they could have been related.

Figure 3-7 also shows that the CR counts were inversely correlated (out of phase) with the altitude of the $0^{o}C$ level most of the time (Note: part of the curves in the graphs is broken because of problems with the counter or the download computer). In the data for mid-September in figure 5, the flat region with a single peak was probably caused by the close approach of a typhoon (Typhoon 娵ork"). The sharp increase of the CR counts may be due to the sudden drop of the atmospheric pressure. However, when the eye of the typhoon passed over our station, the combined effects of cloud shielding and upward shift of the altitude of the $0^{o}C$ level might have resulted in a sharp decrease in the CR counts.

It can be observed that the plots of the CR count and the altitude of the $0^{o}C$ level are positively correlated (in phase) in some period of time. Most of these cases can be attributed to the interference of solar flares, magnetic storms or coronal mass ejection (CME) of the sun. For example, M-Class flare events on 21, 27, 29 August, and 20, 25, 26, 27 October and X-Class events on 2, 31 August might have interfered with the variation pattern near the end of August and October. The solar flares and magnetic storms (often happens after the solar flares) usually suppressed our CR counts as expected.

DISCUSSION

Our observations show that the reversal of CR counts correlates with the atmospheric stability and the sea level CR intensity varies with the altitude of the $0^oC$ level of the atmosphere. These two observations may be attributed to the change of the pressure level height at mid-altitude. According to the ideal gas law, the pressure at the $0^oC$ level height is given by the following equation of the lapse rate, ground temperature and ground level pressure.

$$P_0/P_g = [273/(T_g+273)]^n \qquad (1)$$

where $n = gm/R\theta$. $P_0$ and $P_g$ are the pressure at the $0^oC$ level height and at ground level respectively. $T_g$ is the temperature at the ground level in $^oC$; g is the acceleration of gravity; m is the molar mass of air; R is the ideal gas constant; and $\theta$ is the lapse rate.

From equation (1), it can be seen that an increase in lapse rate will decrease the pressure at $0^oC$ level height. At the same time, an increase in the lapse rate will lower the altitude of the $0^oC$ isotherm level because of an increased vertical temperature gradient (assuming that the ground temperature is the same). That means, an increase in lapse rate will decrease the corresponding altitude of an isobaric level. As the muon intensity profile is determined by the atmospheric depth profile, a decrease in the altitude of isobaric levels will decrease the time for muons to travel to the sea level. This will give a larger CR count at sea level due to the muon decay effect. The mechanism is similar to the temperature effect described in the introduction but the situation now also involves the gradient of temperature. Since the lapse rate is related to the atmospheric stability, the variation of CR counts is then related to the change of atmospheric stability. This is consistent with our observation that a reversal of CR counts pattern occurs whenever a front or trough moves near our station because they are both unstable atmospheric situation.

At the same muon flux at mid-latitude isobaric level (e.g. at 550 mb), an upward shift of the altitude of the $0^oC$ level by 150 m will increase the muon travel time by $5\times 10^{-7}$. If the average energy of the muon is of the order of 1 GeV, the relativistic factor, or, dilation factor $\gamma$ ($\gamma = E/m_0c^2$, where E is the energy of the particle and $m_0$ is the proper mass) will be about 10. Since the lifetime of the muon is 2 $\mu s$ in rest frame, the observed decay lifetime of muon by ground observer will be $\gamma \times 2$ $\mu s$ i.e. $2\times 10^{-5}$ s. The decrease of muon intensity can be estimated by the decay law as $1-\exp(-t/t_0) = 1-\exp(-5\times 10^{-7}/ 2\times 10^{-5}) = 0.025$. Assuming that the pressure value at the altitude of the $0^oC$ level is the same, a 150 m upward shift of the altitude of the $0^oC$ level will decrease the muon intensity by 2.5%. We expect that the actual value will be larger than this estimate because an upward shift of the altitude of the $0^oC$ level also decrease the overall lapse rate. From equation 1, we can see that the pressure at $0^oC$ level will increase. That means the corresponding changes in the altitude of an isobaric level will be larger than the changes in the altitude of the $0^oC$ level. The above estimation is consistent with our presented observation results.

On conclusion, this study shows that the changes of the pattern of the sea level CR counts can be an indication of a change of atmospheric stability. It follows that CR counting may supplement meteorological information in weather forecasting and reporting. We may be able to make use of the CR counts to project or verify the change of atmospheric stability, provided the contributions of the solar effects are quantitatively determined. Further investigation will be needed in order to put this method into practical use.

ACKNOWLEDGEMENT

I would like to express our special thanks to Mr. T.K.Mak for his assistance in setting up of the counting system, data collection, data analysis and valuable discussion on the modification of the system.